# Can Social Networks help the progress of Astrophysics and Cosmology? An experiment in the field of Galaxy Kinematics.

Paolo Salucci

(*SISSA, Via Beirut, Trieste, Italy*)

*Abstract*

This paper is crucial part of an experiment aimed to investigate whether Social Networks can be of help for Astrophysics. In the present case, in helping to eliminate the deep-routed wrong misconception of Flat Rotation Curves of Spiral Galaxies, more rapidly and efficiently than the traditional method of publishing peer-reviewed papers and organizing a number of international conferences. To reach this goal we created the Facebook Group "**Rotation Curve are not Flat** " that we filled with all the evidence necessary for an immediate and definite confrontation with the above fallacious legendary belief. In this paper, we solicit the interested Astrophysicist/Cosmologist FB users to join this group. Finally, the paper informs the Astrophysical Community that a widespread belief is instead an hoax, whose consideration may slow down the progress of science and that must be taken care by innovative means of communicating scientific advances. This test case may anticipate the future in which Web n.0 will become an effective scientific tool for Astrophysics.

e-mail: salucci@sissa.it

## *1. Introduction*

Today social networks (SNs) and Web 2.0 play a major role in many aspects of society, revolutionizing ways of thinking and behaving. Their importance in intra-personal relations is known. Commerce, tourism, games, literature, visual arts, musics, sport, theater, the world of information and that of entertainment, just to name a few, make a large and successfully use of SNs in many different ways. Before proceeding, let us stress that the aim of this paper is not to investigate the sociological/political/philosophical impact of Web 2.0 within society, but to perform, with the help of this Journal, an experiment devised to understand whether SNs can be of help for Astrophysics. Moreover, although we will consider sociological aspects of the scientific community, this paper has no ambition to express something original in this field. First, let us notice that the hard sciences started the Web revolution: it is well known that e-mail and Internet were first widely used in these fields. This tradition in a sense has kept alive; today we have the ArXiv, which has changed the concept itself of submitting a paper to a journal. ADS monitors the impact of the various scientific ideas throughout the community. NED, Hyperleda and similar archives make available to everyone existing observational data. Virtual Observatory has an even more ambitious goal. Again, this is just a sketchy and incomplete account of facts listed to point out that astrophysicists exploit opportunities provided by global networking.

The astrophysical community, like society, is a plural system and it confronts many issues on which there is not a generally agreed truth. Mainstream theories are held by a majority of scientists, but minority views also prosper. Often, very different scenarios, interpretations and discrepant observations all happily coexist. As result of all this, astrophysicists have a non-trivial task in relating with their community and presenting their results by means of traditional methods (publishing papers, attending conferences, discussing with colleagues, etc).

Today, in society people make use of SNs to achieve exactly this goal; we would like to ask if this is also possible in astrophysics, in which web technologies are already present? Blogs and social networks are in fact very effective means for selling new ideas, exposing bad ideas, achieving immediate popularity, building a virtual arena for a global discussion, and storing and retrieving information. A recent political debate highlights the potential for SNs in Science: if they have been used to persuade Americans on the issue of health reform, then could they also be used to convince the astrophysical community to abandon/accept a particular theory?

However, in astrophysics. SNs and blogs do not play an important role in forming the prevalent scientific views, in influencing the scientific policy and in supporting specific ideas. In other words SNs are, currently, a tool of little help for the scientific advance of Astrophysics. At this point, one might be inclined to investigate the causes of this under-utilization, and the prospects for future change, by resorting to statistics and sociology. Instead, we want to take here a more direct route: we devise in this work a test-case experiment to prove or disprove the claim that SNs can be helpful for astrophysics. The experiment goes as follows. First we take into account of the existence of a well-supported observational scenario that, for a number of reasons, has experienced difficulty in completely replacing the previous out-dated view. In other words, within a very well-delimited subject, important observational results are becoming common knowledge at an uncomfortably slow pace. We test our thesis by giving to SNs the mission of rapidly and thoroughly spread the correct scenario throughout the astrophysical community. Our thesis will be affirmed if this mission is accomplished quickly, and with better clarity and visibility and more wide-scale attention and exposure in the community than would be possible with traditional means. The test case is organized in a way that the failure of the mission leaves no hope to SNs to be of any help for astrophysics at the present time.

In the next section we will detail the mission, then we will set up the experiment, a general discussion will be given in the final section.

## 2. *The mission*

The phenomenon of dark matter emerged in the late 1970s from the lack of the expected Keplerian fall-off of circular velocities in the outer regions of spiral galaxies. Soon afterward the misconception of flat rotation curves became popularized, because most of the evidence came from rotation curves (RC) of luminous spirals with small radial variations. In addition, the presumption of perfect flatness of the RCs was incorrectly considered as a consequence of the nature itself of particle dark matter. It was later realized, after the measurement of thousands of high-quality RCs that these rotation curves actually show large variations with radius, and that a similar outcome is predicted within $\Lambda$CDM theory the commonly-accepted scenario of galaxy formation. In the early 90s, the FRC paradigm was abandoned by researchers studying galaxy kinematics, but not entirely by the community at large. These more accurate observations came too late, however, the legend had 15 years to spread among the larger scientific community of cosmologists and extragalactic astrophysicists, and has left in it a strong permanent mark. Of course, several papers describing the correct observational scenario were published, but they were outnumbered by works in which the old paradigm was maintained as a generic and casual statement. As matter of fact, is seems that only a minority of astrophysicists in the past 10 years have searched the literature well enough to find the correct information on the subject, and many contented themselves with this unsubstantiated oral tradition. Moreover, the steadily increasing community is characterized by a high turn-over rate: a significant majority of people entering this field leaves it after a period (~3 years). In many cases, this is too short a time to allow a new scenario to sediment. We discuss this behavior not to theorize on the sociology of astrophysicists, but to explain why standard methods (i.e. publishing papers and reviews, attending topical meetings, and discussing with colleagues) has had only partial success in substituting the misconception of Flat Rotation Curves (FRC) with the correct observational scenario.

## 3. The experiment

This experiment will test the idea that social networks can be an useful tool for the advancement of Astrophysics. In detail, we give to SNs the goal of eliminating the FRC paradigm from Astrophysics. We consider this as a test case, because we choose a very simplified situation. We unquestionably aim to diffuse a truth against a common misconception, which has not even been seriously supported in the literature for quite some time. In the future, if SNs become an important scientific tool, the situation will be much more complicated, as the scientific truth will not necessarily stay ahead of the more innovative side of propaganda.

To reach this goal we will take actions based on reasonable working assumptions. Actions and assumptions will be tested a posteriori. These actions will target the community of astrophysicists, physicists, and mathematicians interested (even loosely) in the issues of dark matter, its alternatives and of galaxy formation and present day properties, which we abbreviate collectively as DMAG.

The first step of the experiment was to create the facebook group "**Rotation Curves are Not Flat**". Its explicit name is intentional: it immediately signals its mission of fostering a change in the paradigm of this particular subject. In the group page members and FB visitors immediately find the proof of the claim: clear cut plots and direct links to crucial published papers/reviews. Furthermore, the group administrators are ready to discuss the topic with members. Possible new papers holding the old paradigm, that might appear in the literature, will be "commented" in the discussion area. Members can upload relevant material to contribute to the discussion. DMAG FB users are asked to join the group but of course do participate to the experiment, independently of their choice.

The second (necessary) step is the publication of this paper to get those DMAG astrophysicists that are not FB users. In fact, this letter a) informs them of the existence of a scientific claim that, for serious reasons, we are attempting to diffuse into the community in an innovative way, b) gives them such evidence in a traditional way (redirecting them to Persic and Salucci 1996, Salucci et al. 2007 and reference therein). Finally they could browse the group pages without joining to it or even to FB. Thus, all people involved in DMAG will be made aware that the idea of flat rotation curves is a common fallacy, and will participate in our experiment, at least by just reading this paper. Let us stress one of the byproducts of SNs: also the more traditional methods benefit of the existence of new methods of spreading scientific results.

The interpretation of the results of the experiment will be simple and clear. Our thesis is proved if on a timescale of months there will be a substantial reduction in the number of submitted papers to astrophysical journals with a FRC statement, (currently, these amount to ~15% of those published in the DMAG subject) or with ambiguous or no statement (~20%), the extinction of papers based on such paradigm (~5%). We also want the corresponding increase of works framed within the correct phenomenology (~60%). Furthermore, we expect a radical shift of thinking also in those many people that, although do not work in the DMAG field, still express FRC convictions in their seminars, lectures, public outreach events and articles, books, reviews and web pages sometime just loosely related to the subject.

The success of the experiment may mark the opening of a new phase in Astrophysics: a truly global community in which results and ideas are really shared and discussed and in which there is less space for prejudices and dogmas.

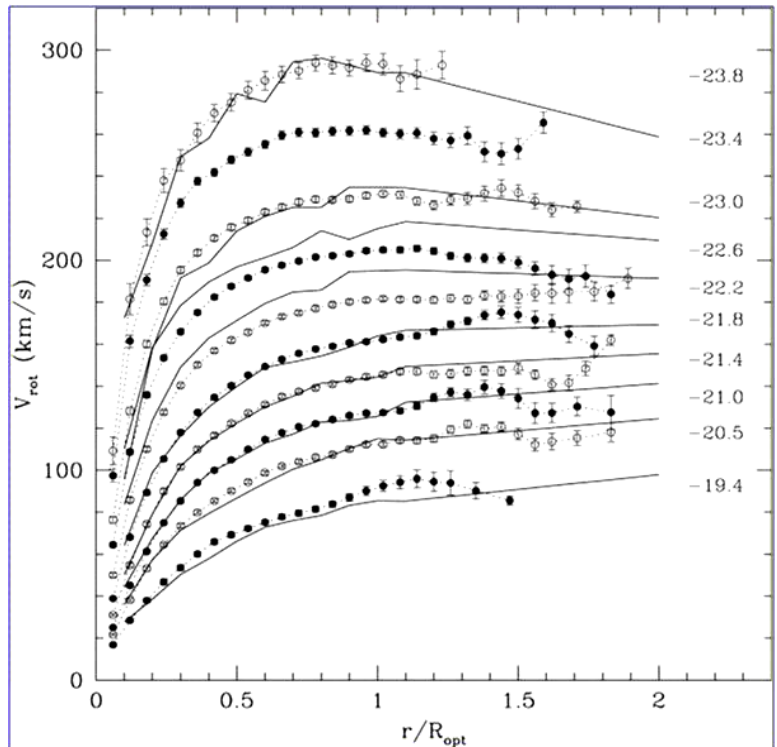

*The actual RC's of Spirals*

However, the experiment will be useful also if it fails in reaching the set goals. In fact, since we will test an extremely clear case where the tasks assigned are not too complicated (in effect they are being tackled with some success by standard methods as well; 5 years ago the above statistics were much worse), if FB fails here, one should seriously wonder about the utility of SNs for the progress of Astrophysics.

## *4. Discussion*

Let us discuss some of the possible criticisms concerning this experiment.
Could astrophysicists reject FB to a level that it would jeopardize the experiment?. First, let us recall again that the group has already 170 members and that the younger generation of scientists are at ease with this technology. Second, FB boosts ideas and accelerate the rate at which these pervade society. A scientist must recognize the difference between a message and the carrier of the message. The experiment may indicate the level of a reluctance to accept SNs in this way present in the community.
Are SNs undermining the peer review system? Why would we want to establish new arenas for scientific debate and who will control them? These concerns are not trivial, however, the effectiveness of this new tool, as it is evident in the ongoing experiment, rests on the fact that the "core message" has already passed a scientific validation. Refereed Journals are and must remain the depository of the current (plural) scientific truth. While the methods with which this is disseminated, discussed, and made universal seem to require modernization.
Why do not just make papers more spectacular and more effective in the propaganda? This is not the real problem. The present system of astrophysical journals is very efficient in promoting the advancement of science, including when this involves new results/ideas, and I would not want scientific publications to lose rigor and reproducibility to become propaganda flyers. The problem is instead that we must find a way to foster the sharing and discussion of new ideas/frameworks within the community of Astrophysics and Cosmology and the interaction itself among scientists in a Community that may be under the risk of becoming to much auto-referential and lazy towards new ideas and paradigm .
A caveat: in this experiment we assume that astrophysicists aim for what they consider to be (or possibly be) a scientific truth and act accordingly. If this assumption is incorrect, the experiment (and astrophysics!) is irrelevant.

## 5. *Conclusion*

Could we now be at the beginning of a true globally interacting astrophysics community? Certainly, today many aspects of astrophysics have global features: the funding, selection of the personnel, methods of working, the topics of research, the interaction with the society, etc. However, in this community there is neither a global link nor a global feedback system that helps/monitors new ideas and new findings in taking their places in the common knowledge of the disclipine. Therefore, we have created an experiment to see whether Social Networks may provide such a link and ask the readers of this paper to participate in it. In detail:

-we ask all scientists studying theoretically, observationally or by computer simulations the fields of Dark Matter, Galaxy Formations, Properties of Galaxies, Alternatives to Dark matter, Detection of Dark Matter, Cosmology and similaria, who are in FB to join the group "**Rotation Curves of Spirals are NOT Flat**"

http://www.facebook.com/home.php?ref=logo#!/group.php?gid=310260450630

and we ask the same kind of scientists who are not on FB, and any astrophysicists interested, to browse the above group, but especially *to seriously consider the claim around which such group has been built* and, if it is the case, to look at the supporting evidence in the papers provided below.
This paper sets the experiment, the results will be published in due time.

*Acknowledgments*

P.S. thanks R. Gilmore for a very useful discussion.